\documentclass[twocolumn,showkeys,amsmath,amssymb,floatfix,pre]{revtex4-1}
\usepackage{graphicx}
\usepackage{mathcomp}

\newcommand{\be}{\begin{equation}}
\newcommand{\ee}{\end{equation}}

\newcommand{\bea}{\begin{eqnarray}}
\newcommand{\eea}{\end{eqnarray}}

\begin{document}

\preprint{APS/PRE}

\title{Thermodynamic and dynamic anomalies in a one dimensional \\ lattice model of liquid water}
\author{Marco Aur\'elio A. Barbosa}
\thanks{Corresponding author}
\affiliation{Faculdade UnB Planaltina, Universidade de Bras\'ilia, Planaltina-DF, Brazil}
\author{Fernando Vito Barbosa}
\affiliation{Instituto de F\'isica, Universidade de Bras\'ilia, Bras\'ilia-DF, Brazil}
\author{Fernando Albuquerque de Oliveira}
\affiliation{Instituto de F\'isica, Universidade de Bras\'ilia, Bras\'ilia-DF, Brazil}

\date{\today}

\begin{abstract}
We investigate the occurrence of waterlike thermodynamic and dynamic anomalous behavior in a one dimensional lattice gas model. 
The system thermodynamics is obtained using the transfer matrix technique and anomalies on density and thermodynamic response functions are found. When the hydrogen bond (molecules separated by holes) is more attractive than the van der Waals interaction (molecules in contact) a transition between two fluid structures is found at null temperature and high pressure. This transition is analogous to a `critical point' and intimately connects the anomalies in density and in thermodynamic response functions. Monte Carlo simulations were performed in the neighbourhood of this transition and used to calculate the self diffusion constant, which increases with density as in liquid water. 
\end{abstract}


\maketitle

\section{Introduction}

Liquid water has many properties which are recognized as being anomalous when compared to other nonbonded liquids with the same molecular size.
As an example, the isothermal compressibility and the constant pressure heat capacity present minimum as a function of temperature at ambient pressure. In addition, the thermal expansion coefficient is negative below $4\tccentigrade$, indicating that density  increases anomalously with temperature~\cite{debenedetti04:review,Franks:water:matrix}. Besides thermodynamic anomalies, water present dynamic anomalies. For temperatures below $283\tccentigrade$ there is a region where the self-diffusion constant increases as a function of pressure~\cite{debenedetti04:review,marcia:jcp}.

The thermodynamic anomalies seems to be inter-related and different thermodynamic scenarios were proposed to describe these relations~\cite{debenedetti04:review}. Here we will focus our discussion on the second critical point scenario~\cite{poole92:nat}, that is supported by the model studied here (as will be shown latter). According to this scenario, the diverging thermodynamic behavior of the response functions is associated with the critical point arising from metastable liquid-liquid phase transition that occurs in the supercooled regime, in an experimentally unaccessible temperature below the homogeneous nucleation temperature. This scenario was originally observed and proposed by means of computer simulations of atomistic models for liquid water~\cite{poole92:nat}, but there are indirect experimental evidences for the existence of a liquid-liquid phase transition in supercooled water on amorphous\cite{mishima98:nat1,mishima98:nat2} and confined water\cite{chen08:epj}.

The relation between thermodynamic and dynamic anomalies have been subject of discussion in the literature~\cite{debenedetti01:nat,chen08:epj,Truskett07:jcp1,truskett07:jpcb}. Using atomistic models of water, Errington and Debenedetti found that the anomalously diffusive region surrounds the region of density anomaly in the temperature vs. density plane~\cite{debenedetti01:nat}. In addition, measures of translational and rotational structure also reveals unexpected behavior, that could be connected to the anomalies on density and diffusion~\cite{debenedetti01:nat}. These relations, usually called \textit{hierarchy of anomalies}, are not restricted to water models and where observed in computer simulations of $\mathtt{SiO_2}$~\cite{sharma06:jcp,charusita09:prer}, $\mathtt{BeF_2}$~\cite{charusita09:prer}, and core-softened models of fluids~\cite{marcia08:jcp,truskett06:jcp1,truskett07:jpcb}.

With the aim of investigating the relation between thermodynamic and dynamic anomalies of water we propose a one dimensional lattice model of water. Lattice and off-lattice one dimensional models of water were proposed by Ben-Naim\cite{ben-naim08:jcp,ben-naim08:jcp2,Ben-Naim:statmech}, Bell\cite{bell69:jmath}, and others~\cite{robinson96:prl,stanley99:pre}, to investigate the so-called thermodynamic anomalies and even the unusual solvation behavior presented by water. Nevertheless, and as far as we know, the behavior of the equilibrium diffusion constant of these models were not investigated. 

The thermodynamics of our model is obtained exactly using transfer matrix technique and its dynamics is investigated through Monte Carlo simulations. A waterlike anomalous behavior is found both on density and self diffusion constant, as observed in water~\cite{debenedetti04:review} and atomistic models of water~\cite{marcia:jcp,stanley07:pre:yan}.

This paper is organized as follows. In section~\ref{model} we build the model's Hamiltonian and investigate its ground state. In section~\ref{thermo} the thermodynamics of the model is analysed and discussed. Monte Carlo simulations are used to calculate the self-diffusion constant on section~\ref{simulation} 
and the relation between thermodynamics and kynetics is explored on section~\ref{connection}. Final remarks are made on the last section.

\section{\label{model}The model}

The model consists of a linear lattice whose sites can be either occupied by molecules or empty. Two interactions are defined: a short-range van der Waals attraction between nearest neighbours and a hydrogen bond between second nearest neighbours separated by holes. An occupation variable $\eta_k$ is assigned to each site $k$ to indicate the presence ($\eta_k=1$) or the absence ($\eta_k=0)$ of a molecule. With these definitions the effective Hamiltonian in the grand canonical ensemble is written as:
\bea
\mathcal  H & = & - \sum_i \left [ \epsilon_{vdw} \eta_i \eta_{i+1} + \epsilon_{hb}\eta_i( 1-\eta_{i+1})\eta_{i+2} \right ] \nonumber \\ 
            &   &  - \mu \sum_i  \eta_i, \label{eq:h}
\eea
where $\epsilon_{vdw}>0$ defines the strength of the van der Waals attraction, $\epsilon_{hb}>0$ the strength of the hydrogen bond, and $\mu$ is the chemical potential. 

In order to obtain a preliminary insight about the model we investigate the ground state by looking at three states putative states at null temperature: a gas (G),  a bonded fluid (BF) and a dense fluid (DF). The gas phase is an empty lattice which `coexists' with the fluid phase of lower free energy at null pressure. 
The bonded fluid is a half filled and fully bonded lattice, while the dense fluid is a filled lattice without bonds. For a lattice with size $L$ and periodic boundary conditions, the grand canonical free energies of the fluid phases are $\Phi_{BF}=-(\epsilon_{hb}+\mu)L/2$ and $\Phi_{DF}=-(\epsilon_{vdw}+\mu)L$.
We will only consider parameters that correspond to the hydrogen bond being more attractive than the van der Waals interaction, i.e., $\epsilon_{hb} < \epsilon_{vdw}$, in consistence with real water.  In fact, this condition also ensures the predominance of a bonded fluid at low temperatures and pressures in the range $0 < P < P_c$, with $P_c = \epsilon_{hb
} - \epsilon_{vdw}$. At null temperature, a transition between the BF and the DF happens at $P_c$, as observed by Sadr-Lahijany \textit{et al.} in a similar continous model~\cite{stanley99:pre}. Here we follow ref.~\cite{stanley99:pre} on calling this null temperature transition as the `second critical point', in addition to the `critical point' found at $T=0$ and $P=0$. Further discussions about the `second critical point' of this system will be made on section~\ref{lowt}.

\section{\label{thermo}Thermodynamics}

\begin{figure}
\begin{center}
\includegraphics[scale=0.65]{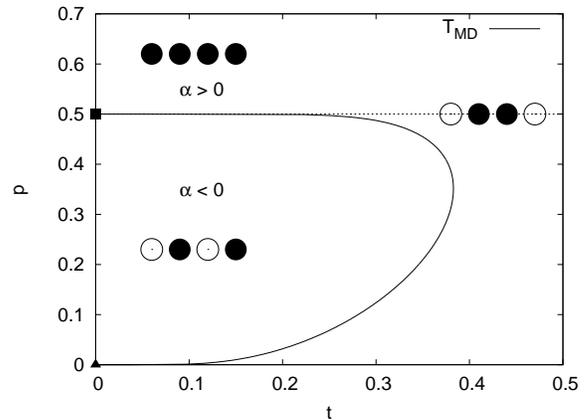}
\caption{Temperature of maximum density (TMD) in the $p=Pl/\epsilon_{vdw}$ vs. $t=kT/\epsilon_{vdw}$ phase diagram, for $\epsilon_{hb}/\epsilon_{vdw}=3/2$. Note that the TMD starts in the low pressure `critical point' at ($0$,$0$) and ends in the high temperature `second critical point' at ($0$,$0.5$).}
\label{fig:tmd}
\end{center}
\end{figure}

We use the transfer matrix technique to exactly calculate the grand canonical partition function (more details can be found on appendix~\ref{tmt}) of Hamiltonian~(\ref{eq:h}) as
\be
\Xi (T,L,\mu) = \lambda^L = e^{\beta P L}, \label{eq:pf}
\ee
where $T$, $L$, $\mu$ and $P$ are, respectively, the thermodynamic variables temperature, system's length, chemical potential and pressure, and $\beta=1/kT$, with $k$ the Boltzmann constant. From $\Xi (T,L,\mu)$ the density $\rho$ becomes:
\be
\rho = \frac{ (\lambda-1)\left[ (a \lambda+b) (\lambda-1) +1  \right ]   }{ (\lambda-1)^2(a\lambda+2b)+3\lambda-2}.\label{eq:rho}
\ee
where $a=e^{\beta \epsilon_{vdw}}$ and $b=e^{\beta \epsilon_{hb}}$. It is shown on appendix A that the density can be written in terms of these quantities as:

In addition to the density $\rho$, the densities (per site) of hydrogen bonds ($\rho_{hb}$) and nearest neighbours ($\rho_{nn}$) are also important to relate  the liquid structure to the anomalous behavior and are calculated on appendix A. The mean number of hydrogen bonds ($n_{hb}$) and nearest neighbours ($n_{nn}$) per particle can be calculated from those expressions as $n_{hb} = \rho_{hb} / \rho $ and  $ n_{nn} = \rho_{nn} / \rho$.

The thermodynamic response functions are important to characterize waterlike behavior, particularly because they indicate the possibility of a critical behavior on the supercooled regime (as is expected to occur in liquid water). Here we investigate the thermal expansion coefficient ($\alpha$) and isothermal compressibility ($\kappa$), defined as:
\begin{subequations}
\bea
\alpha & = & -\frac{1}{\rho} \left (\frac{\partial \rho}{ \partial \ T} \right )_P, \\
\kappa & = & \frac{1}{\rho} \left (\frac{\partial \rho}{ \partial \ P} \right )_T.
\eea
\end{subequations}
Expressions for these functions are lengthy and they will not be presented here.

\subsection{\label{lowt}Low temperature limit}

\begin{table}
\caption{\label{tab:nulltemp}Value of various quantities while approaching the transition at $(T_c=0 , P_c= \epsilon_{hb}-\epsilon_{vdw})$ through different limits. Only the temperature dependence is presented for $\alpha$ and $k_T$.}
\begin{ruledtabular}
\begin{tabular}{cccc}
		& $P \rightarrow P_c^-$ & $P=P_c$ 	& $P \rightarrow P_c^+$ \\ \hline
$  \rho$  	&        $1/2$ 		&   	$2/3$	&	$1$	 	\\ 
 $\rho_{hb}$	&         $1/2$		&   	$1/3$	&	$0$	 	\\ 
 $\rho_{nn}$	&        $0$ 		&   	$1/2$	&	$1$		 \\ 
$\alpha \sim$	&	$-1/T$	&   	$0$ 	& $ 1/T$ 	\\ 
$k_T \sim$		&         $ 1/T$	&   $1/T$	&	$1/T$	 \\ 
\end{tabular}	
\end{ruledtabular}
\end{table}

Let us note that for $P=0$ expression~(\ref{eq:rho}) results in a null density $\rho=0$. Nevertheless, at null temperature the limiting value of the density depends on how the point ($T=0$,$P=0$) is approached, due the `transition' occuring at this point. While approached from above, the density is compatible with the BF ground state (see the second column of Tab.~\ref{tab:nulltemp}). In addition, both $\alpha$ and $k_T$ diverges while approaching this `critical point' (not shown). An equivalent behavior was previously obtained by Sadr-Lahijany \textit{et al.} in a similar continous model~\cite{stanley99:pre}. Even more interesting is the low-temperature behaviour of the system in the neighbourhood of the second critical point ($T=0$,$P=P_c$). 
In what follows we analyse this behavior focusing on density $\rho$ and its derivative $\alpha$. From now on we assume $P>0$. Using this one obtains the low temperature limit of eq.~(\ref{eq:rho}) as
\be
\rho \approx \frac{a\lambda +b}{a\lambda+2b}, \label{eq:rholowt}
\ee
where it was assumed that $\lambda \gg 1$ at small temperatures. By applying this expression to the point $(T=0,P=0)$ it follows that $\rho \rightarrow 1/2$ while the point is approached \textit{from above}, as discussed before. 

Now, we note that $\lambda_c = e^{\beta P_c} = b/a$ in the neighbourhood of the fluid-fluid transition at \[T=T_c=0, \;\;\;P = P_c = \epsilon_{hb} - \epsilon_{vdw} \] 
and define 
\[
P = P_c + \Delta, \;\;\; \lambda = \lambda_c \delta, \;\;\; \delta = e^{\beta \Delta},
\]
in order to simplify eq.~(\ref{eq:rholowt}). While approaching this transition it can be found that $\delta \rightarrow 0$ for $\Delta>0$,   $\delta =1$ for $\Delta=0$ and $\delta^{-1} \rightarrow 0$ when $\Delta<0$. Using these definitions, eq.~(\ref{eq:rholowt}) is written as
\be
\rho = \frac{1+\delta}{2+\delta}.\label{eq:rhoscp}
\ee
The low-temperature limits of the density and the derivatives of~(\ref{eq:rhoscp}) are obtained in a straightforward manner (see Tab.~\ref{tab:nulltemp}). 
It is particularly interesting to note that the values expected from section~\ref{model} are reobtained when the `critical' pressure is approached from below or from above, see table~\ref{tab:nulltemp}. Nevertheless, when null temperature limit is approached with $P=P_c$ one finds $\rho \rightarrow 2/3$, which is different from both BF and DF. This is an indication of a new liquid structure, which can be represented through a serie of two neighbouring water molecules surrounded by empty sites. By comparing the expected grand canonical free energies, it was verified \textit{a posteriori} that this state coexists with both BF and DF exactly at $P_c$. The existence of such a hidden state make the analysis of the ground state limit more complex. Nevertheless, a simple derivation similar to the one presented above, was used for obtaining the low-temperature limiting values of the densities of hydrogen bonds~(\ref{eqap:nhb}) and nearest neighbours~(\ref{eqap:nnn}) presented on table~\ref{tab:nulltemp}.

We finish this analysis noting that the behavior of the thermal expansion coefficient in the neighbourhood of the fluid-fluid phase transition changes qualitatively depending on how the `second critical point' is approached. For a linear approach $\Delta = kT \tan \theta$ we find $\alpha \propto \tan \theta/T$, which is reasonable since the density anomalous increase with temperature in region dominated by BF fluid structure, for $\Delta <0$ or equivalently $P<P_c$.

\subsection{Liquid structure}

In what follows we analyse the thermodynamic liquid structure in the pressure vs. temperature phase diagram for the parameter $\epsilon_{hb}/\epsilon_{vdw}=3/2$. Similar features are found for other parameter values, provided that $\epsilon_{vdw}<\epsilon_{hb}$.

Fig.~\ref{fig:tmd} shows that the line of temperature of maximum density (TMD) starts at $T=0$ exactly at the BF-DF transition, which has the peculiar property that \[\alpha(0,P_c) = \lim_{T \rightarrow 0, P=P_c }  \alpha(T,P) \rightarrow 0.\] This line reaches its maximum temperature at $t=kT/\epsilon_{vdw} \approx 0.388(1)$ and then its temperature decreases with pressure until reaching the G-BF transition located at $(T_c=0,P_c=0)$. As discussed previously by Sadr-Lahijany \textit{et al.}~\cite{stanley99:pre}, the G-BF transition is a remanent of a gas-liquid phase transition of a system with higher dimensionality, and can be found in simple 1D fluids with attractive short-range interactions. The very fact that the TMD connects two transitions (G-BF and BF-DF) seems to be a peculiar property of some 1D models, since a similar feature also occurs in the 1D lattice model proposed by Bell~\cite{bell69:jmath} and for some parameters in reference~\cite{stanley99:pre}, but not int the continuous models studied in~\cite{bell69:jmath,ben-naim08:jcp}.

\begin{figure*}
\begin{center}
a \includegraphics[scale=0.35]{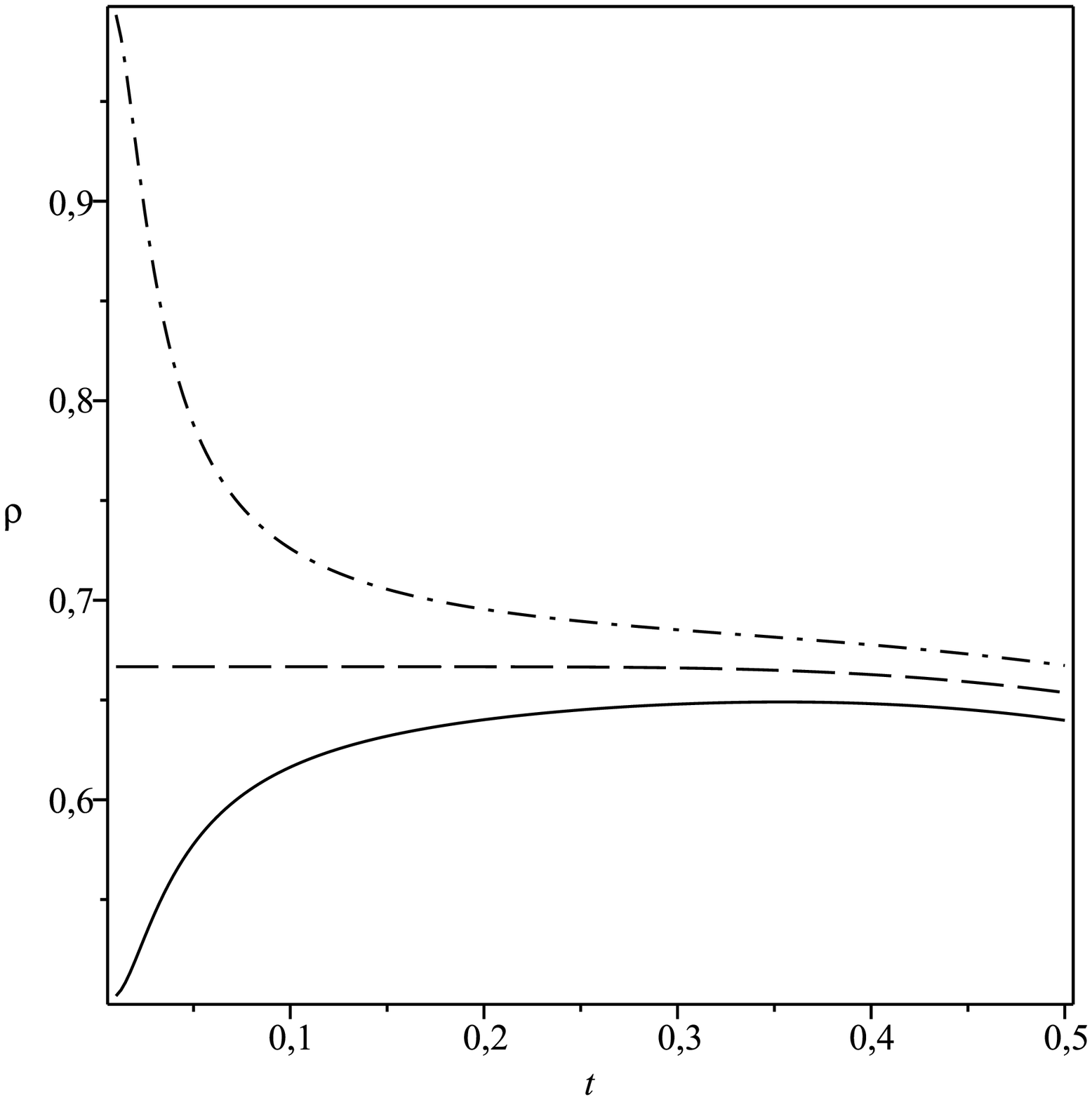} b\includegraphics[scale=0.35]{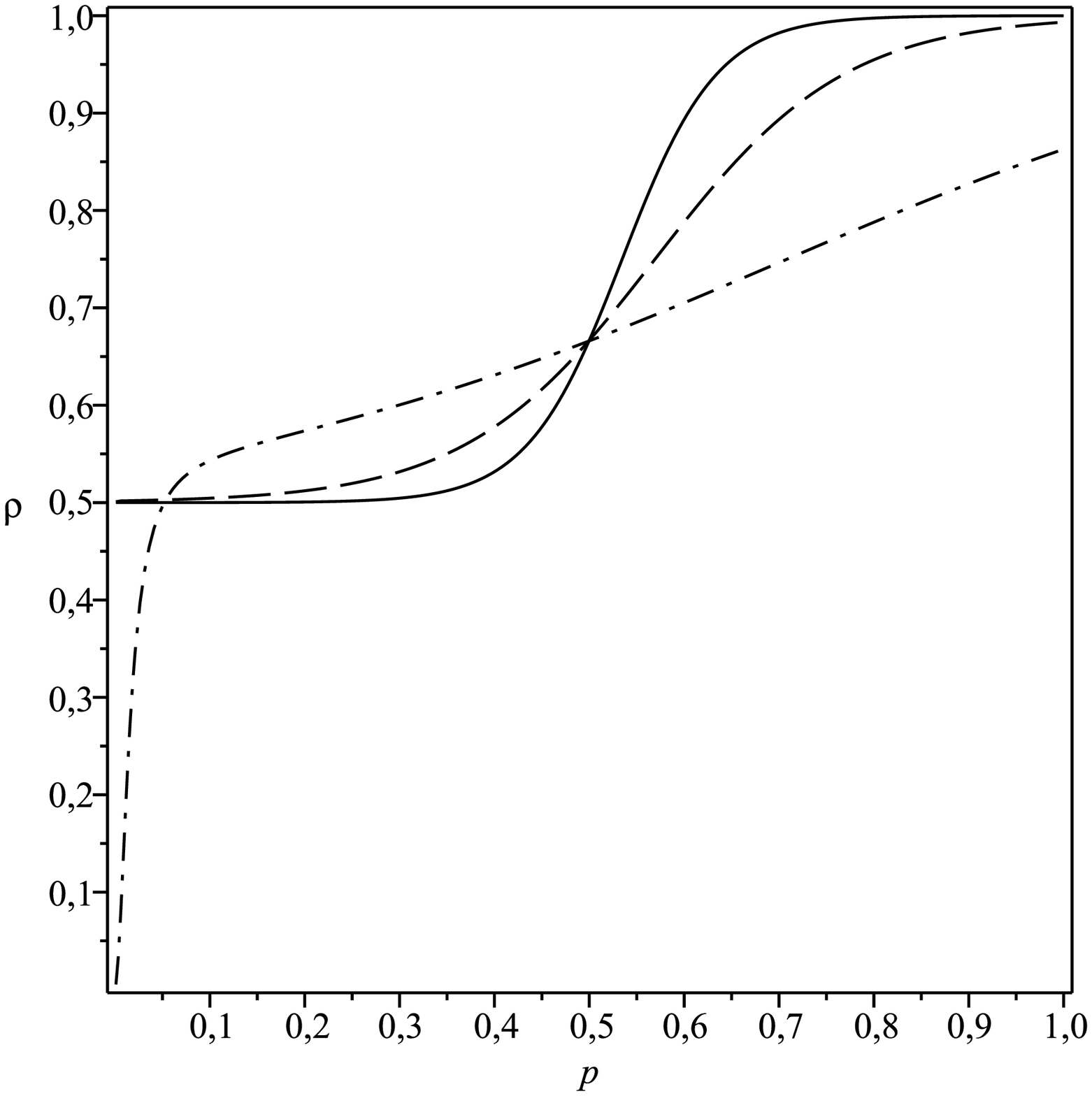}  \\
c\includegraphics[scale=0.35]{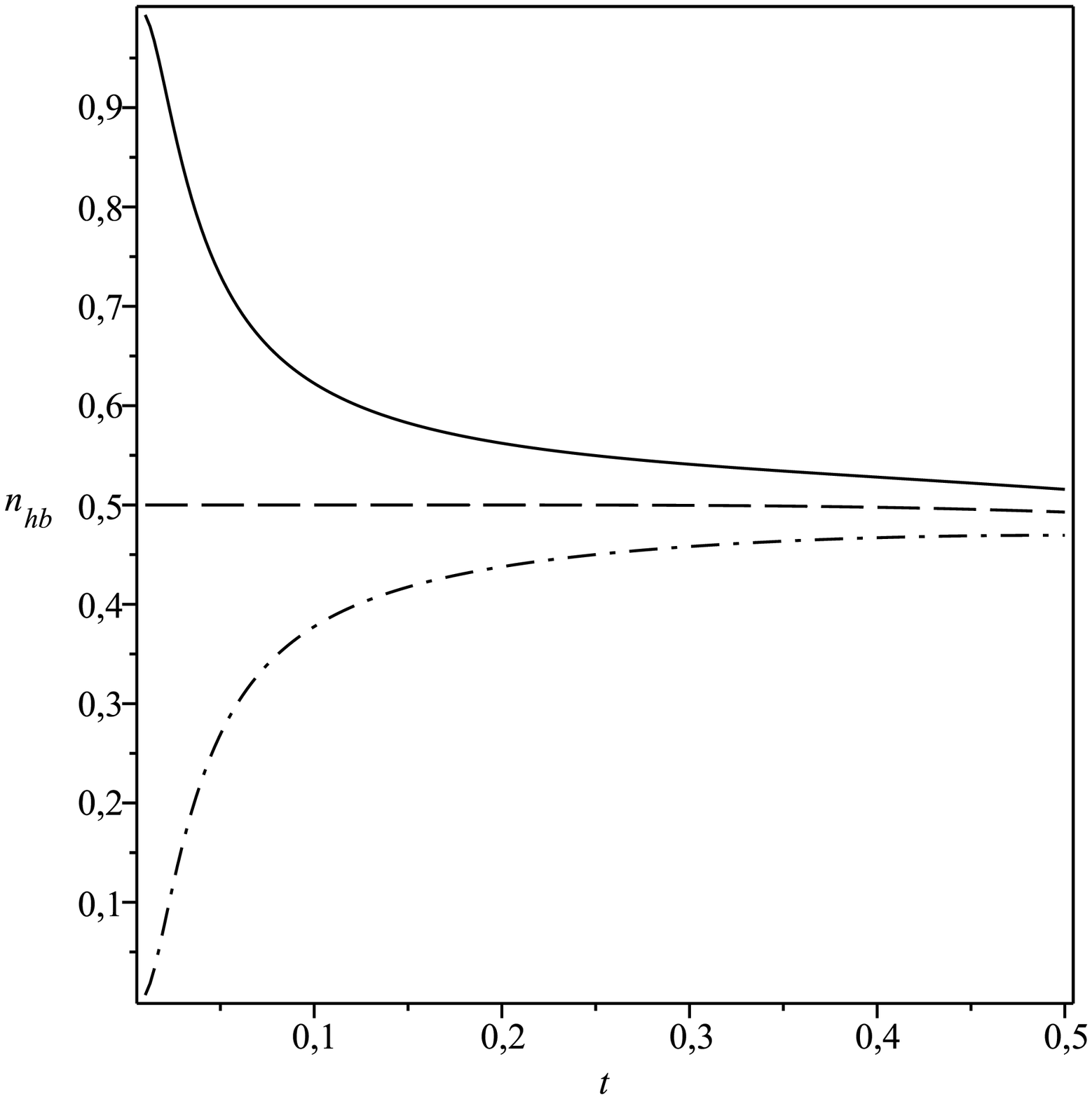} d\includegraphics[scale=0.35]{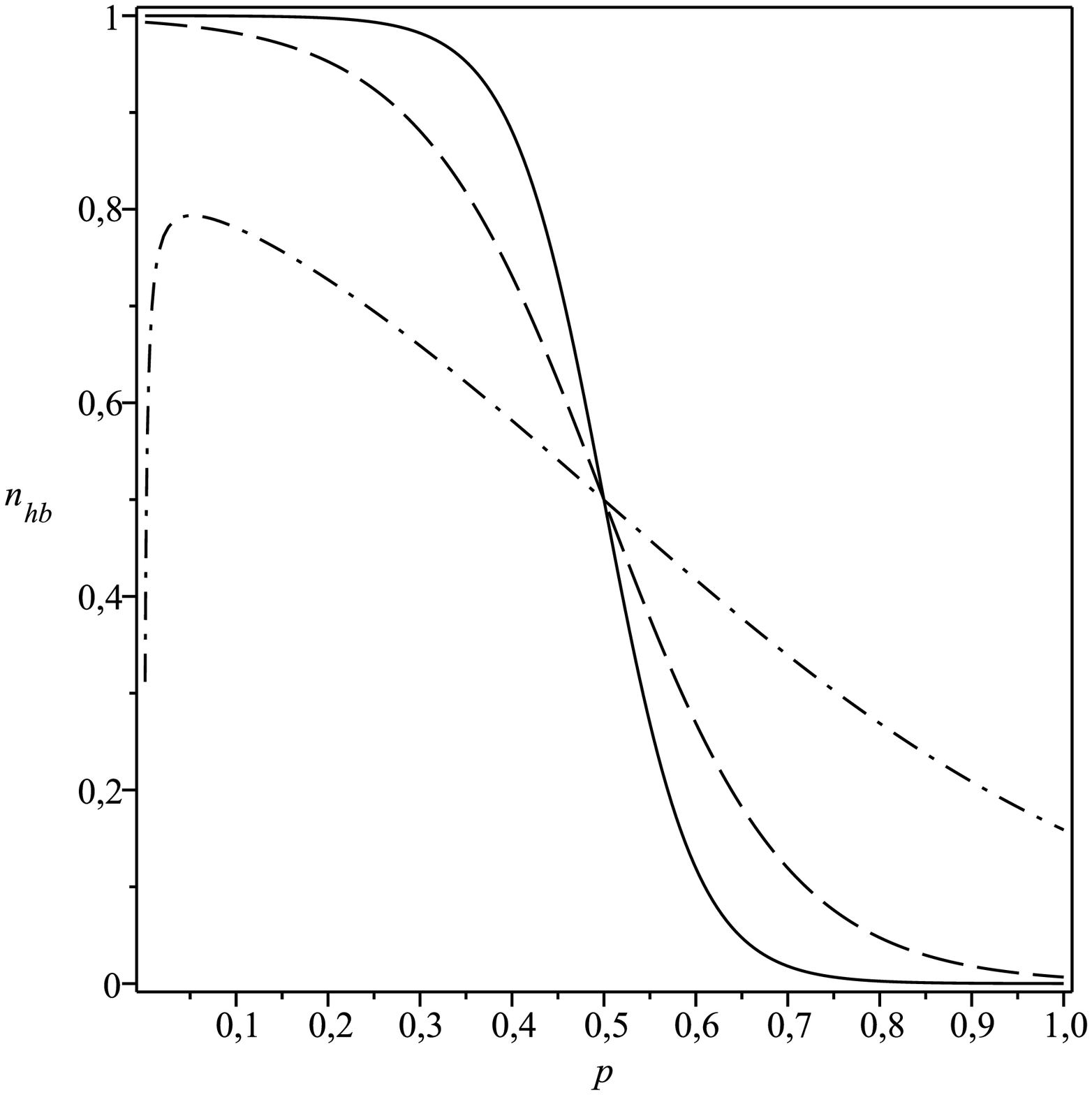}  \\
e\includegraphics[scale=0.35]{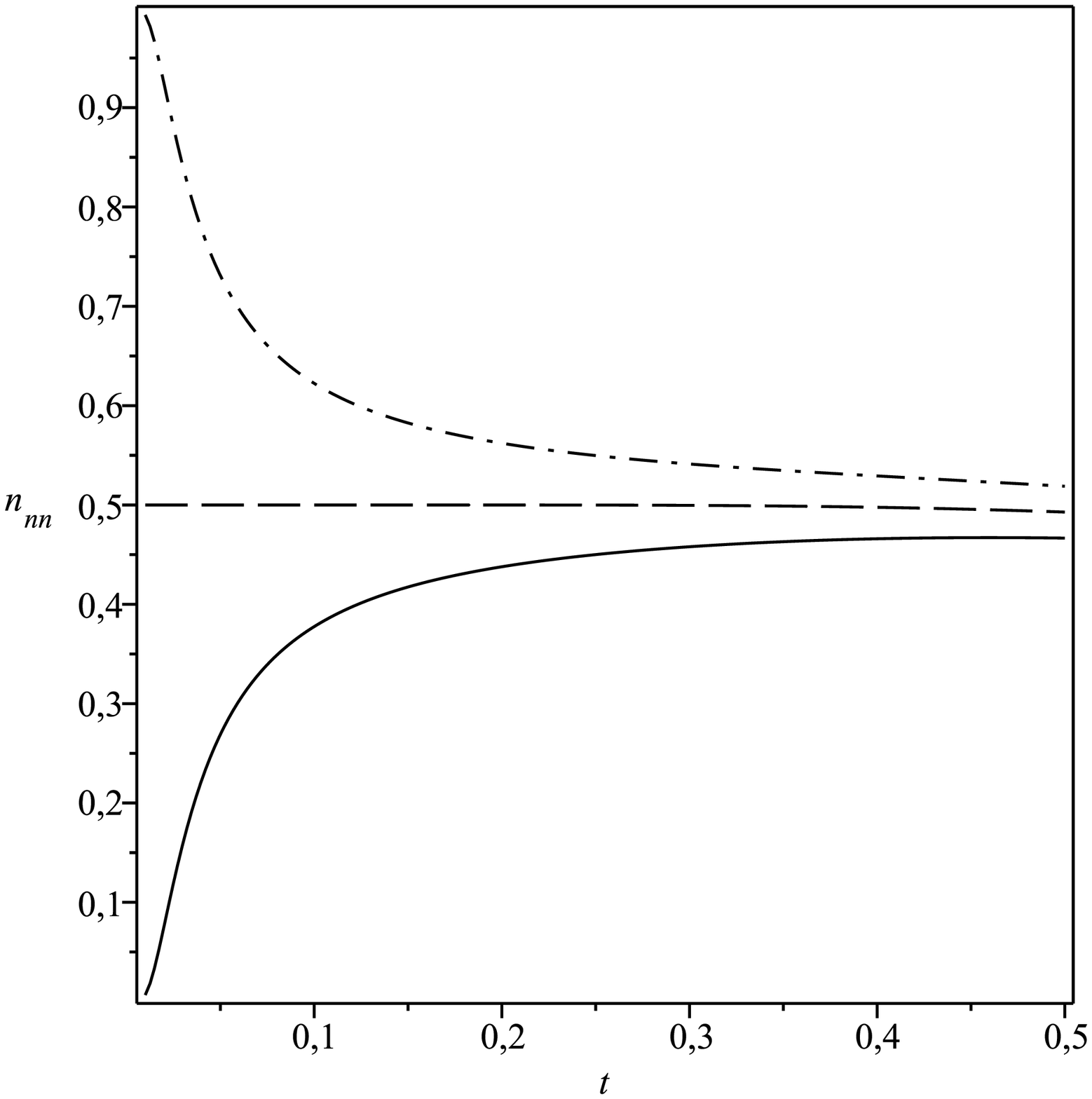} f\includegraphics[scale=0.35]{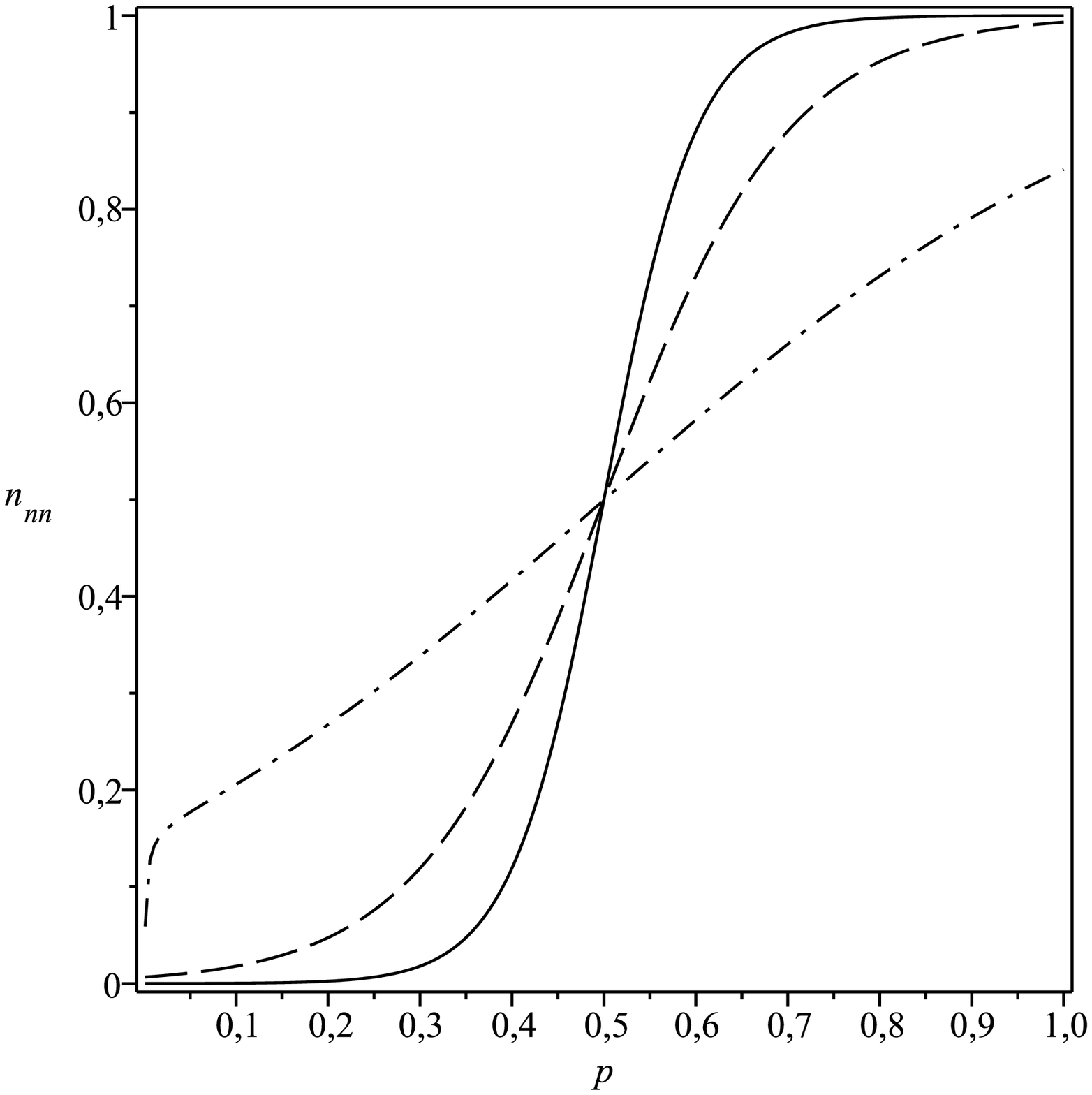}  \\
\caption{The temperature dependence of $\rho$ (a), $n_{hb}$ and $n_{nn}$ is shown on the left side for $p=0.45$ (dashdot), $0.50$ (dash) and $0.55$ (solid). The pressure dependence of the same quantities is shown on right side for $t=0.05$ (solid), $0.1$ (dash) and $0.3$ (dashdot). Units are the same used in Fig.~\ref{fig:tmd}.}
\label{fig:dens}
\end{center}
\end{figure*}

In Fig.~\ref{fig:dens} the behavior of density, number of hydrogen bonds  and number of nearest neighbours per particle is compared as a function of temperature and pressure. Figs.~\ref{fig:dens}~(a), (c) and (e) show that the anomalous increase in density with temperature (for $P<P_c$) is followed by a decrease on the number of hydrogen bonds and by an increase on the number of first nearest neighbours. For $P=P_c$, not only $\alpha \approx 	0$ is persistent for a wide range of temperatures, but $(\partial n_{hb}/\partial T)_P \approx (\partial n_{nn}/\partial T)_P \approx 0$ is also found in the same region. Figs.~\ref{fig:dens}(b), (d) and (f) show that the continuos transition between BF and DF progressively becomes more abrupt as the temperature is lowered. From these figures, it is also evident  that $\rho$, $n_{hb}$ and $n_{nn}$ are essentialy not changing with temperature at $P=P_c$, since different isotherms cross at this point. These results indicate that the intermediate fluid structure discussed in the previous section is well populated in a borderline separating a BF-like region from a DF-like region. It is possible that this population occurs because the intermediate structure is an unavoidable intermediate step for changing between the BF and DF liquid structures. A last comment about Figs.~\ref{fig:dens} (b), (d) and (f) is approapriate at this point: it is evident from the high temperature curves that a fast transition between the G and BF structures takes place at low pressures. This is associated with the low-$T$ and low-$P$ part of the TMD, which happens with a fast but small increase in density with temperature, as also observed in Figs.~5 and 6 from reference~\cite{bell69:jmath}.

\begin{figure}
\begin{center}
\includegraphics[scale=0.35]{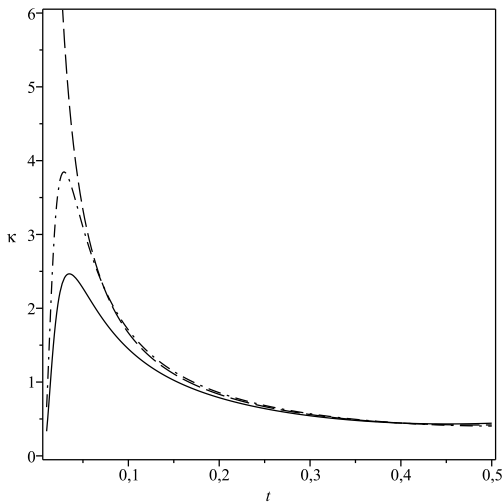}
\caption{Isothermal compressibility $\kappa$ as a function of temperature at same pressures as in Fig.~\ref{fig:dens}: $p=0.45$ (solid), $0.50$ (dashed) and $0.55$ (dashdot).}
\label{fig:kappat}
\end{center}
\end{figure}

\begin{figure}
\begin{center}
\includegraphics[scale=0.35]{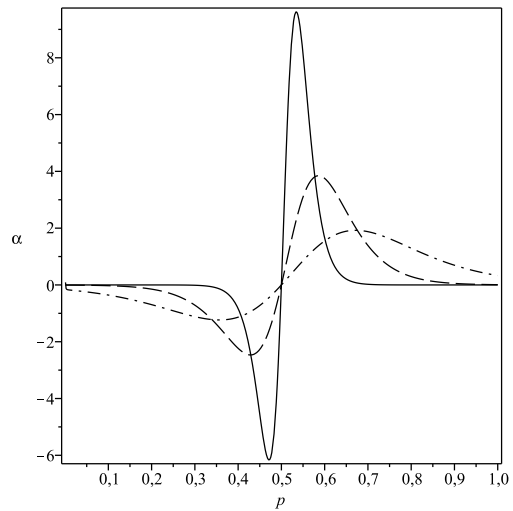}
\caption{Thermal expansion coefficient $\alpha$ as a function of pressure at temperatures $t=0.02$ (solid), $0.05$ (dashed) and $0.1$ (dashdot).}
\label{fig:alpha}
\end{center}
\end{figure}

The isothermal compressibility $\kappa$ is presented in Fig.~\ref{fig:kappat} as a function of temperature, at the same pressures as Figs.~\ref{fig:dens}(a), (c) and (e). Note that $\kappa$ increases with decreasing temperature in the neighbourhood of of the BF-DF transition, and that it diverges when approaching this transition with $P=P_c$. The diverging behavior of $\alpha$ can be more easily understood as a function of pressure with fixed temperature, as in Fig.~\ref{fig:alpha}. Note that, while approaching the BF-DF transition through a pressure decrease, $\alpha$ is diverging to $-\infty$  in the BF side and to $+\infty$ on the DF side, as discussed in the previous section. 

In the next section we turn our attention to Monte Carlo simulations used to study the anomalous behavior of the diffusion constant of this model.

\section{\label{simulation}Monte Carlo Simulations}

\begin{figure}
\begin{center}
\includegraphics[scale=0.65]{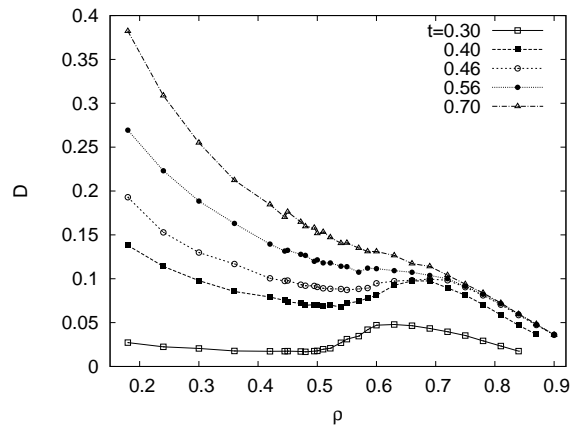}
\caption{Self-Diffusion constant as a function of density at fixed temperature.}
\label{fig:diffusion}
\end{center}
\end{figure}

Monte Carlo simulations were performed in the canonical ensemble using jumps to nearest and next-nearest neighbour sites. Next-nearest neighbour jumps are necessary to observe diffusion in one dimensional lattice fluids since it is impossible for the particles to \textit{turn around} their neighbours~\footnote{Without next nearest neighbour jumps this model presents a more complex subdiffusive dynamics.}.

Each simulation starts with $N$ particles placed randomly along $V$ sites of a linear lattice with periodic boundary contourn, at temperature $t$. The initial configuration is recorded in a vector $\vec{X}(0)$ with the position of each molecule. On each MC timestep $N$ particles are randomly selected to jump and, in the absence of a collision, the Metropolis algorithm is applied to test the acceptance of the movement~\cite{binder}. 
After an equilibration time $\tau_{eq}$ the mean square displacements in relation to the initial configuration is calculated and recorded at every $\tau_{r}$ timesteps until the simulation time $\tau_f$. A number $\mathcal R$ of different simulations is performed to avoid dependences on initial conditions. The diffusion constant is then obtained by interpolating the Einstein equation:
\be
\left \langle \sum_{k=1}^{N} \left [ X_k(\tau) - X_k(0) \right ]^2  \right \rangle_\mathcal{R} = 2 D (t,\rho) \tau
\ee
where the average is performed over the $\mathcal R$ initial conditions.

\begin{figure}
\begin{center}
\includegraphics[scale=0.7]{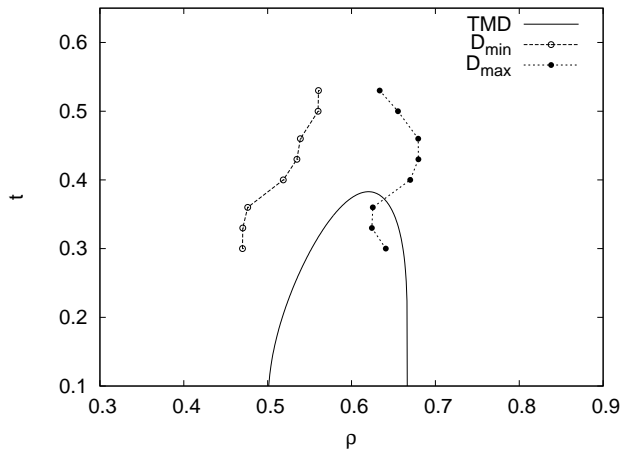}
\caption{The location of the maximum (filled circles) and minimum (empty circles) values of the self-diffusion constant are 
compared to the line of temperature of maximum density (line) in the $t$ vs. $\rho$ phase diagram.}
\label{fig:hierarchy}
\end{center}
\end{figure}

This protocol was used to simulate a system of size $V=1000$, with the temperature $t$ ranging from $0.30$ to $0.90$ and the density in the interval $0.18 \leq \rho \leq 0.9$. The simulation times used for this system was $\tau_{f} = 10^7$, $\tau_{eq} = 1.2 \times 10^6$ and $\tau_{r} = 5\times 10^3$; and a number $\mathcal R = 10$ of different initial configurations were performed on each point. The error on the diffusion constant was smaller than $0.5\%$ in all cases.

\subsection{\label{connection}Anomalous Diffusion and Hierarchy of anomalies}

The self-diffusion is shown in Fig.~\ref{fig:diffusion} as a function of density for five different temperatures. At high temperatures, and particularly for $t=0.7$ in Fig.~\ref{fig:diffusion}, the diffusion decreases with density, as expected in a system with simple hard core particles. For temperatures below $t \approx 0.55$ it is possible to observe points of minimum ($D_{min}$) and maximum ($D_{max}$) diffusion, as a function of density. Between $D_{min}$ and $D_{max}$ the self-diffusion constant anomalously increase with density, as happens for water in a region of the phase diagram~\cite{debenedetti04:review}. The presence of a minimum in diffusion was observed in atomistic~\cite{debenedetti01:nat} models of liquid water, as well as continuous~\cite{marcia08:jcp,barraz09:jcp} and lattice~\cite{marcia07:pa,marcia07:pa2,szortyka10:jcp} models with waterlike behavior but, as far as we know, this is the first time that this behavior is observed in a one dimensional lattice model.

In order to locate $D_{min}$ and $D_{max}$ points on the neighbourhood of minimum and maximum diffusion were fitted with the function:
\be
\label{eq:adjust}
D(t,\rho) = \left ( a\rho^3+b\rho^2+c\rho+d \right ) \exp \left(  e\rho^2+f\rho+g \right ),
\ee
with $a$, $b$, $c$, $d$,  $e$, $f$, and $g$ as numerically adjustable parameters. At low temperatures ($t \leq 0.43$), two parameter sets were used to fit $D(t, \rho)$, with at least $10$ points around each extrem. For temperatures higher than $t = 0.43$ a single function was used to fit at least $17$ points covering both extrems. Maximum and minimum values of diffusion were calculated through the derivative of eq.~(\ref{eq:adjust}). Points of extremum diffusion are compared to the exact TMD line in the density vs. temperature phase diagram on Fig.~\ref{fig:hierarchy}. 
Even though not reaching the typically lower temperatures of the region anomalous in density, the simulations indicate that for less dense states ($\rho \lesssim 0.6$) the line of minimum diffusion occurs at temperatures higher than the TMD line. For higher densities, the line of maximum diffusion presents a reentrance and it crosses the TMD line, indicating that an strict \textit{hierarchy of anomalies} is absent on this simplified model (at least for $\epsilon_{vdw}/\epsilon_{hb}=2/3$). 

By comparing Fig.~\ref{fig:hierarchy} and similar results from other lattice models~\cite{marcia07:pa,marcia07:pa2,szortyka10:jcp} one observes an overall tendency of overlap between regions of anomalous diffusion and density. Nevertheless, the so-called hierarchy of anomalies does not always occur in simple hierarchical structure, as found in atomistic models of water~\cite{debenedetti01:nat,stanley07:pre:yan}. In two-dimensional lattice models with waterlike behavior a hierarchy of anomalies was found in a reversed order, i.e., with the TMD line covering the anomalously diffusive region~\cite{marcia07:pa2,szortyka10:jcp}. On the other side, a three dimensional lattice model also presents a line of extremum diffusion crossing the TMD, as in this work, but with the region of diffusion anomaly covering the TMD at low densities~\cite{marcia07:pa}. Considering these models and the results presented
on this paper it is possible  to conclude that there is an overall tendency for the concomitant ocurrence of density and diffusion anomalies but the existence of an hierarchical order between these anomalies depends on dimentionality and even on possibily on the detailed nature of the interactions between molecules.

\section{\label{end}Conclusion}

We introduced a simple one dimentional lattice-gas model that reproduces waterlike anomalies on its thermodynamics and dynamics through transfer matrix technique and Monte Carlo simulations. Although other models were used to investigate water's peculiar properties\cite{bell69:jmath,ben-naim08:jcp,stanley99:pre,robinson96:prl}, (as far as we know) this is the first time this type of model is used to investigate the inter-relations between thermodynamic and dynamic waterlike anomalies.

Our results show that the region of thermodynamic anomalies is located in the neighbourhood of the transition between two structured fluids: a bonded fluid (BF) and a dense fluid (DF). The null temperature `second critical point' is linked to this transition and the relation between the criticality and the temperature of maximum density can be understood from the behavior of the thermal expansion coefficient in the vicinity of the critical point.

An increase on the diffusion coefficient with density (at fixed temperature) was also found in the neighbourhood of BF-DF transition. Even though existing an overall tendency for the concomitant ocurrence of the regions of anomalous diffusion and density, a strict \textit{hierarchy of anomalies} is absent on this model.

As final remark we mention that the simplicity of the current model makes it an ideal prototype for more detailed investigations on the inter-relations between waterlike thermodynamic and dynamic anomalies. We are now working on analytical approaches to the diffusion constant of this model using techniques from non-equilibrium statistical mechanical. Another approach that is currently under investigation involves the calculation of the diffusion constant through the Kubo relation to understand the connection between memory and diffusion in more complex topologies~\cite{morgado02:prl}.

\section*{Acknowledgements}
We acknowledge B. Widom for usefull discussions and C. E. Fiore dos Santos for carefully reading the original manuscript of this paper. 
This work has been supported by the CNPq, FAPDF and FINATEC. MAAB acknowledges finantial support 
from the National Science Foundation via grant CHE-0842022 during part of this work.
\appendix 

\section{\label{tmt}Transfer matrix technique}

Here we use the transfer matrix technique to calculate the grand canonical partition function of linear chain with Hamiltonian~(\ref{eq:h}).
More detailed survey of the technique can be found on textbooks of statistical mechanical~\cite{salinas:introduction,Ben-Naim:statmech}.
Since our model does present interactions up to three neighbouring lattice sites, we will rewrite the Hamiltonian as
\be
\mathcal{H} = \sum_i h(\eta_{i-1},\eta_{i},\eta_{i+1}),
\ee
where periodic boundary conditions are assumed implicitly. The three-site interaction Hamiltonian $h(\eta_{i-1},\eta_{i},\eta_{i+1})$ is defined as
\bea
\label{eq:h3}
h(\eta_{i-1},\eta_{i},\eta_{i+1}) & = &- \frac{1}{2}  \epsilon_{vdw} \left ( \eta_{i-1} \eta_{i} + \eta_i \eta_{i+1} \right ) \nonumber \\
                                  &   & - \epsilon_{hb}\eta_{i-1}( 1-\eta_{i})\eta_{i+1} -\mu \eta_i.
\eea

A simple procedure to represent the transfer matrix of in a one dimensional system with first and second nearest neighbour interactions is described by Ben-Naim~\cite{Ben-Naim:statmech}. Following this procedure, the grand canonical partition function can be written as 
\be
\Xi (T,L,\mu) =\sum_{\vec{\eta}} \prod_i  Q (\eta_{i-1},\eta_{i},\eta_{i+1}), \label{eqap:pf}
\ee
where the sum is made over all lattice states $\vec{\eta} \equiv \{ \eta_0, \ldots, \eta_{L-1} \}$ and $Q (\eta_{i-1},\eta_{i},\eta_{i+1}) \equiv e^{-\beta h(\eta_{i-1},\eta_{i},\eta_{i+1}) }$. Let us now define an auxiliar occupation variable $\eta'_i$ and a new Boltzmann weight as
\be
P (\eta_{i-1},\eta_{i},\eta'_i,\eta_{i+1}) = \left \{ \begin{array}{cl}
                                                      Q (\eta_{i-1},\eta_{i},\eta_{i+1}) &, \eta_i = \eta'_i \\
						      0 &, \eta_i \neq \eta'_i.
                                                   \end{array} \right.
 \ee
It is not difficult to use $P (\eta_{i-1},\eta_{i},\eta'_i,\eta_{i+1})$ to write (\ref{eqap:pf}) as trace of a matrix:
\be
\Xi (T,L,\mu) = \texttt{Tr} \left \{ \mathcal{P}^L \right \},\label{eqap:trace}
\ee
where the elements of matrix $\mathcal{P}$ are ordered through binary numbers whose \textit{bits} are occupation variables, and are given by:
\[ \left [  \mathcal{P} \right ]_{ \eta \eta',\eta''\eta''' } = P ( \eta ,\eta',\eta'',\eta'''). \]

In the thermodynamic limit, only the largest eigenvalue ($\lambda$) of $\mathcal P$ contributes to the right hand side of eq.~(\ref{eqap:trace}), and this equation becomes
\be \Xi = \lambda^L = \exp(\beta P L ),\label{eqap:p}\ee
where the characteristic polynomial of $\mathcal{P}$ is 
\be
\label{eqap:characteristic}
\lambda^3 - (1+za)\lambda^2+z(a-b)\lambda + z(b-1)= 0.
\ee

By solving together equations~(\ref{eqap:p}) and~(\ref{eqap:characteristic}) one can obtain either $P\equiv P(T,\mu)$ or $\mu \equiv \mu(T,P)$. Here we follow Bell in ref.~\cite{bell69:jmath}, who noted that $z=e^{\beta \mu}$  can be rearranged from~(\ref{eqap:characteristic}) as a function of $\lambda$, $a$ and $b$; leaving the possibility of  calculating the density as
%\begin{subequations}
\bea
\rho & = & \frac{z}{\lambda L} \left( \frac{\partial \ln \Xi  }{ \partial z} \right )_{a,b} \label{eqap:rho0}, \nonumber \\
     & = &  \frac{z}{\lambda} \left [ \left ( \frac{\partial z}{\partial \lambda} \right )_{a,b}  \right ]^{-1}, \nonumber \\ 
     & = &  \frac{ (\lambda-1)\left[ (a \lambda+b) (\lambda-1) +1  \right ]   }{ (\lambda-1)^2(a\lambda+2b)+3\lambda-2} \label{eqap:rho1}.
\eea
%\end{subequations}

Finally, the same procedure can be used to calculate the densities of hydrogen bonds and nearest neighbours (per site) as
\begin{widetext}
\bea
\rho_{hb} & = & \frac{b(\lambda-1)}{  \lambda \left [  3\lambda^2-2(1+za)\lambda+z(a-b) \right ]},\label{eqap:nhb} \\
\rho_{nn} & = & \frac{za}{ \lambda \left \{  1+bz/\lambda^2+z(2\lambda-1)[\lambda (\lambda-1)]^{-2} \right \} } \label{eqap:nnn}.
\eea 
\end{widetext}

%\bibliography{/home/aurelio/Dropbox/papers/references/references,/home/aurelio/Dropbox/papers/references/books}

%

%\pagebreak

\end{document}